\documentclass[showpacs, preprintnumbers, nofootinbib, aps, prd, superscriptaddress,10pt, showkeys, notitlepage, twocolumn]{revtex4-1}

\usepackage{graphicx,amssymb,amsmath,amsthm,amsfonts,epsfig}

\usepackage[linktocpage,breaklinks]{hyperref}
\usepackage[usenames,dvipsnames]{color}
\usepackage{epstopdf}
\usepackage{aas_macros}
\definecolor{darkred}{rgb}{0.5,0,0}
\definecolor{darkgreen}{rgb}{0,0.5,0}
\definecolor{darkblue}{rgb}{0,0,0.5}
\definecolor{prussian}{rgb}{0.0, 0.19, 0.33}
\definecolor{richelectricblue}{rgb}{0.03, 0.57, 0.82}
\definecolor{teal}{rgb}{0.0, 0.5, 0.5}
\definecolor{mediumseagreen}{rgb}{0.24, 0.7, 0.44}
\definecolor{lust}{rgb}{0.9, 0.13, 0.13}
\definecolor{ballblue}{rgb}{0.13, 0.67, 0.8}
\definecolor{darkcyan}{rgb}{0.0, 0.55, 0.55}
\definecolor{mountainmeadow}{rgb}{0.19, 0.73, 0.56}
\definecolor{palecarmine}{rgb}{0.69, 0.25, 0.21}
\definecolor{richcarmine}{rgb}{0.84, 0.0, 0.25}
\definecolor{tangelo}{rgb}{0.98, 0.3, 0.0}
\definecolor{venetian}{rgb}{0.784,0.031,0.082}
\definecolor{bdfrance}{rgb}{0.192,0.549,0.906}

\hypersetup{colorlinks=true, citecolor=venetian,
linkcolor=bdfrance, urlcolor=lust}
\usepackage{amsmath,amssymb}
\usepackage{mathtools}
\usepackage{amsbsy}
\usepackage{bm}
\usepackage{float}

\def\apjs{{The Astrophysical Journal Supplement}}
\def\apjl{{ApJL}}

\def\aap{{A\&A}}

\def\04a{{2004 a}}
\def\04b{{2004 b}}

\newcommand{\be}{\begin{equation}}
\newcommand{\ee}{\end{equation}}
\newcommand{\bea}{\begin{eqnarray}}
\newcommand{\eea}{\end{eqnarray}}

\newcommand{\ta}{\tau_{\rm amb}}
\newcommand{\Ug}{U_{\text{grav}}}
\newcommand{\Um}{U_{\text{mag}}}
\newcommand{\Ur}{U_{\text{rot}}}
\newcommand{\Mr}{M_{\text{rot}}}

\newcommand{\Ms}{M}

\newcommand{\Rs}{R}
\newcommand{\Mtov}{M_\text{{\tiny TOV}}}

\begin{document}

\title{Magnetically supramassive neutron stars}

\begin{abstract}
\noindent{It is commonly believed that neutron stars exceeding the maximum mass limit for stability could be formed in the 
aftermath of binary neutron star mergers, enjoying a short life of metastability before losing centrifugal support and collapsing 
to a black hole. It is suggested here that a similar scenario could take place when the remnant's excess mass is supported by 
an ultra-strong $(\gtrsim 10^{17}\,\mbox{G})$ magnetic field that could be generated during, and shortly after, coalescence. 
We show that such `magnetically supramassive' neutron stars could stave off collapse and survive for a few years before their 
magnetic energy is sufficiently dissipated due to ambipolar diffusion. In addition, we speculate on multi-messenger signatures 
of such objects and discuss the robustness of our results against limitations placed by neutron superfluidity and magneto-thermal evolution.}
\end{abstract}

\author{Arthur G. Suvorov}
\email{arthur.suvorov@manlyastrophysics.org}
\affiliation{Manly Astrophysics, 15/41-42 East Esplanade, Manly, NSW 2095, Australia}
\affiliation{Theoretical Astrophysics, Eberhard Karls University of T{\"u}bingen, T{\"u}bingen, D-72076, Germany}

\author{Kostas Glampedakis}
\email{kostas@um.es}
\affiliation{Departamento de F\'isica, Universidad de Murcia, Murcia, E-30100, Spain}
\affiliation{Theoretical Astrophysics, Eberhard Karls University of T{\"u}bingen, T{\"u}bingen, D-72076, Germany}

\date{\today}
 
\maketitle



\section{Introduction}
Determining the chemical makeup that defines the equation of state (EOS) of neutron star matter constitutes one of the key 
open problems in high-energy astrophysics. Matching data from electromagnetic, and more recently gravitational-wave (GW), observations of extreme 
phenomena, such as short gamma-ray bursts (SGRBs), with theoretical predictions from general-relativistic magnetohydrodynamics (GRMHD) offers an 
invaluable tool in this respect. As evidenced by the joint GW-GRB event GW170817 detected by Fermi and the advanced Laser Interferometer GW Observatory 
(aLIGO) \cite{ligo17,ligo18}, neutron star mergers can be production sites for SGRBs, the prompt-emission and afterglow light-curves of which reveal unique 
information about the nature of the remnant \cite{rowl13,ravi14,lg16,suvk21}. 




If the merging stars are not too massive, a third, more extreme neutron star may emerge from the crash site rather than a black hole.
It is generally posited that this star can have one of three fates depending on its mass, $\Ms$, in relation to the \emph{maximum} mass 
(for a given EOS) resulting from the integration of the Tolman-Oppenheimer-Volkoff (TOV) stellar structure equations, $\Mtov$: 
(i) Long-term stability, where the star survives indefinitely, for $\Ms \leq \Mtov$;
(ii) medium-term metastability for $\Mtov < \Ms \lesssim 1.2 \Mtov$, where uniform rotation stabilises the remnant (often termed `supramassive'; \cite{sterg03}), or 
(iii) short-term metastability for $1.2 \Mtov \lesssim \Ms \lesssim 1.6 \Mtov$, where differential rotation stabilises the remnant (`hypermassive'; \cite{weih18}). 
In this Letter we suggest that the remnant neutron star has a \emph{fourth} option, namely, long-term metastability, where the collapse 
is eventually instigated by core magnetic field decay. Such neutron stars could be fittingly called `\emph{magnetically supramassive}'.

In particular, the magnetar subclass of neutron stars may contain extremely strong magnetic fields within their stellar cores, which could potentially help 
stabilise them against gravitational collapse \cite{card01,sterg03,dex17}. Violent dynamo activity at birth \cite{td96}, 
possibly in combination with 
the Kelvin-Helmholtz \cite{ciolfi20} and magneto-rotational \cite{reb21} instabilities, may thus facilitate the growth of magnetic pressures within the 
remnant to the point that the birth mass may non-negligibly exceed $\Mtov$, even without rapid rotation. Following a swift $(\gtrsim$ seconds) spin-down 
phase, the star would then survive on a magnetic-diffusion timescale on the order of $\sim 1-10$ years, 
depending on the core temperature, internal field 
strength, birth mass, and EOS \cite{gr92,ho12}. Here we provide some analytic estimates for mass limits and collapse times of magnetically supramassive 
stars based on magneto-thermal arguments, finally offering some discussion on observational signatures of such objects, most notably from SGRBs.



\section{Maximum mass of magnetic stars} 
In much the same way that rotational kinetic energy can help stave off gravitational collapse, 
so too can magnetic energy. The Lorentz force associated with the magnetic field contributes an effectively anisotropic stratification
which, for poloidal fields, drives the star towards an oblate shape and can work together with the hydrostatic pressure to resist gravity \cite{card01,dex17}. 
Microphysical effects, such as Landau quantization and the spin polarization of neutrons within the stellar core, also start to influence the bulk properties 
of the star for ultra-strong magnetic fields \cite{sedrakian17}. Overall, there is a secondary stiffening effect on the EOS for super-Schwinger fields \cite{brod00}, and even more 
massive stars can be produced. A rigorous calculation of the maximum mass sustained by a neutron star under the influence of rotation and/or a strong 
magnetic field requires the numerical integration of the GRMHD structure equations for a given realistic EOS. Fortunately, for the purposes
of this work, this complication can be avoided and we can instead rely on a far simpler analytical approach based on energy arguments.  
  
Consider first the classic, rotationally supramassive case. The maximum mass of a static configuration is $\Mtov$, though rotation contributes to the 
available energy pool and pushes this limit higher. Assuming a uniformly rotating star, the sum of the gravitational and rotational kinetic energies are 
$\Ug + \Ur = -3 G \Ms^2 / 5 \lambda \Rs +  \tfrac{1}{2}I_0 \Omega^2$, for moment of inertia $I_{0} \approx \tfrac{2}{5} M R^2$ and rotational velocity $\Omega$, 
where we have introduced the phenomenological parameter $\lambda$ to account for EOS and GR effects (for a uniform Newtonian model, $\lambda=1$). 
The \emph{maximum} mass of the rotating configuration, $\Mr$, can then be estimated by considering a star rotating at the mass-shedding (Kepler) limit, 
$\Omega_{\rm K} \approx \sqrt{G \Mtov / \Rs^3}$, and {equating the sum of the kinetic and potential energies of this more massive star with the 
maximum binding energy available to a static star.} Solving $-3 G \Mtov^2/ 5 \lambda \Rs = -3 G \Mr^2/ 5 \lambda \Rs + \tfrac{1}{2} I_{0} \Omega_{\rm K}^2$, 
we find $\Mr = \Mtov  \sqrt{1 + \tfrac{\lambda}{3}}$, which is in remarkably good agreement with the numerical simulations for $\lambda \lesssim 2$ \cite{sterg03,weih18}.

The above procedure can be similarly carried out with the magnetic energy, $\Um = \tfrac{1}{6} B^2 R^3$, in place of the rotational energy. 
Note, however, that $B$ here is \emph{not} the surface field strength but rather a volume-averaged internal field strength, 
the magnitude of which may be dominated by the outer-core toroidal field or inner-core poloidal field. As such, even if a relatively conservative surface field is realised, $B_{\text{surf}} \lesssim 10^{16} \text{ G}$, the value of $B$ here could potentially approach the 
Virial limit, $B_{\rm max} \sim 10^{18} \lambda^{-1/2} (M/M_{\odot})(R/10 \text{km})^{-2}$~G, depending on the field topology. 
{We emphasise however that it is unclear whether fields of this strength are ever reached in Nature; even in merger simulations 
with large $(\sim 10^{15}$~G) seed fields, magnetic energies tend to saturate at a few times $10^{51}$ erg \cite{ciolfi20,shib21}, implying an upper limit $B_{\text{max}} \sim 6 \times 10^{16} (U_{\text{mag}}/ 2 \times 10^{51} \text{ erg})^{1/2}(R/15 \text{km})^{-3/2}$~G. 

However, intense and largely unresolvable magnetic substructures are prevalent in many studies, and it is conceivable that greater amplifications could be attained if finer spatial grids, necessary to fully resolve the Kelvin-Helmholtz and/or magneto-rotational instabilities, 
are employed (see Ref. \cite{ciolfi20} for a thorough discussion). On the observational side, there is reasonable evidence that at least some of the X-ray afterglows seen to follow many SGRBs are powered by spindown energy injections from a newborn magnetar \citep{ravi14,lg16,suvk21}. In many cases, fittings within this paradigm favour polar fields $B_{p}$, again likely lower than the internal field strength, that are a few by $10^{16} \text{ G}$; in some rare instances though, most notably GRB 100625A, best-fit values of $B_{p} \gtrsim 10^{17}~\text{G}$ have been reported \citep{rowl13}. These estimates however assume perfectly efficient emissions, and therefore represent upper limits.}

\begin{figure}[htb!]
\begin{center}
\includegraphics[width=0.42\textwidth]{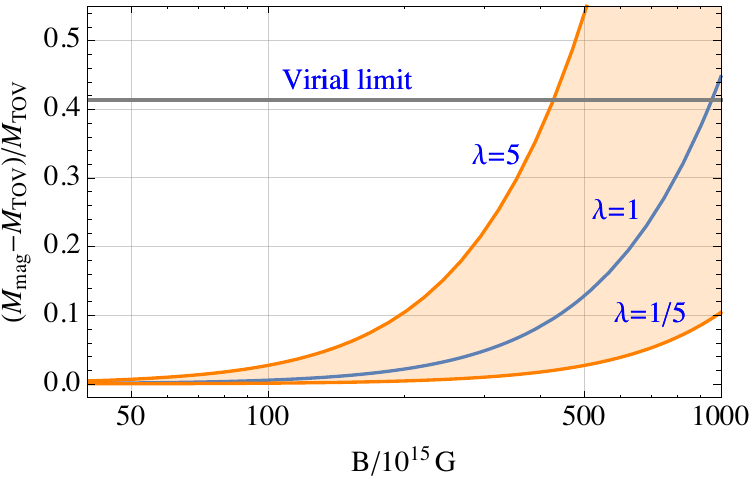}
\end{center}
\caption{Maximum mass of a magnetic star as a function of the volume-averaged magnetic field $B$, 
estimated through Eq.~\eqref{eq:magmass}, for $\Mtov=2.2 M_\odot$ and $R = 15\,\mbox{km}$.
The shaded region $(0.2 <\lambda<5)$ should capture most of the uncertainty related to the EOS and GR gravity (see main text).
The horizontal line represents the maximum mass set by the Virial limit, $M_{\rm max}/\Mtov \approx 1.414 $.}
\label{fig:Mmag}
\end{figure}

Either way, for the maximum magnetic mass we estimate
\be
\frac{M_{\rm mag}}{\Mtov} =  \left ( 1 + \frac{5 \lambda}{18}\frac{ B^2 R^4}{G \Mtov^2} \right )^{1/2}.
 \label{eq:magmass}
\ee
This result is again in reasonable agreement with the numerical calculations. For instance, for a $n=1$ polytropic EOS, Ref. \cite{card01} found that the 
(baryonic) mass increases from $2.19 M_{\odot}$ to $2.28 M_{\odot}$ (i.e., a $4\%$ increase) for a GR stellar model with radius $R = 13.8$ km 
and magnetic dipole moment $\mu \approx B R^3 / (2 \sqrt{g_{rr}})= 2 \times 10^{35} \text{ G cm}^{3}$ (for Schwarzschild factor $g_{rr}$). 
This result, and others for the same EOS, matches well with the simple formula \eqref{eq:magmass} for $\lambda \lesssim 3$. For the APR EOS \cite{apr98} 
(which passes constraints coming from GW170817 \cite{ligo18}), we find instead $\lambda \sim 1$, unless $B \gtrsim 10^{18}$ G in which case the aforementioned 
stiffening results in a better-fit value $\lambda \gtrsim 2$  \cite{card01}. Similar results are found using simulation data from other works for various EOS, 
such as Ref. \cite{dex17}. We therefore consider the range $0.2 \leq \lambda \leq 5$ for demonstration purposes; Figure \ref{fig:Mmag} illustrates 
the formula \eqref{eq:magmass} in this respect.
%

\section{Magnetic field decay and collapse}
A magnetic field residing in the interior of a neutron star can dissipate via the main mechanisms of
Ohmic decay, likely accelerated by Hall drift, and ambipolar diffusion~\cite{gr92}. For the magnetic field-temperature
parameter space relevant to the core of a newborn neutron star, the field decay is dominated by ambipolar diffusion.
This mechanism involves the drift of the charged particles (protons and electrons) relative to the 
neutron fluid. The magnetic field, anchored to the charged fluids, follows this motion and the 
induction-generated electric field leads to magnetic flux transport and field line straightening. The drift culminates in 
the release of magnetic energy while heating up the star \cite{gr92,ho12}.

The characteristic decay time for the magnetic field reads $\ta = L / w_{\rm amb}$, where $L$ denotes a 
typical distance over which the magnetic field varies and $w_{\rm amb}$ is the typical velocity lag between
the charged and uncharged fluids. This lag is determined by the balance between the Lorentz force and
the inter-particle collisional forces. 
Typically $\ta$ is broken up into solenoidal and irrotational components associated with the Helmholtz-Hodge 
decomposition of $\mathbf{w}_{\rm amb}$, though such a distinction is not necessary for our simple demonstration. 
The MHD equations associated with the system imply that~\cite{gr92}
\be
\ta \approx 25\, L_{5}^2\, B_{17}^{-2} \,T_{9}^{2} \,\left( \rho / \rho_{\text{nuc}} \right)^{2/3}  \text{ yr},
\label{tamb}
\ee
where $L_5 = L/10^5\,\mbox{cm}$, $B_{17} = B/10^{17}\,\mbox{G}$, and $T_9 = T/10^9\,\mbox{K}$. 
(Note, however, that since $|\textbf{B}|/|\nabla \textbf{B}| \propto \ell^{-1}$ for a pure $\ell$-pole, $L$ may be \emph{lower} in the early stages as 
the field unknots from a highly-tangled configuration, as would be expected in a newborn magnetar \cite{td96,reb21}, thereby accelerating the decay.) 

\begin{figure}
\begin{center}
\includegraphics[width=0.42\textwidth]{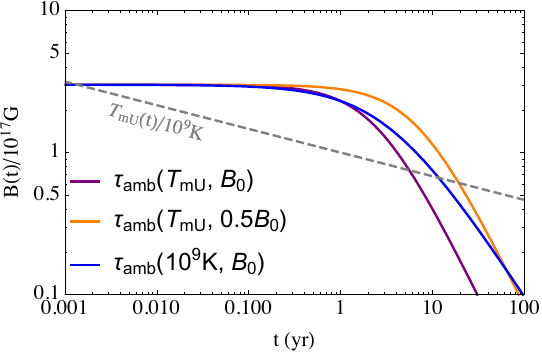}
\end{center}
\caption{Time evolution of the magnetic field (solid curves) due to ambipolar diffusion, according to Eq.~\eqref{eq:magfield},
for different choices for $\ta (T,B)$; see Eq.~\eqref{tamb}. Overplotted for reference is the mUrca temperature $T_{\rm mU}$ (dashed curve) from Eq.~\eqref{eq:TmU}.}
\label{fig:Bevol}
\end{figure}

A rigorous treatment of the ambipolar diffusion-driven decay of the magnetic field involves the numerical evolution of the system's 
coupled GRMHD-thermal equations. However, and in spite of recent progress~\cite{pass17}, such calculation has not been completed 
yet. For the approximate analysis of this paper it is sufficient to work with the phenomenological evolution law~\cite{ho12}
\be
B(t) = B_0 \left ( 1 + \frac{t}{\ta} \right )^{-1}.
\label{eq:magfield}
\ee
The ambipolar timescale, $\ta$, can be treated as a constant with fixed values of $B$ and $T$ during the magnetic
field evolution or it can be promoted to a `dynamical' parameter with a time-varying  temperature $T(t)$. 
At densities $\rho\approx \rho_{\rm nuc}$, the stellar core is expected to cool via neutrino emission produced by the 
modified Urca reactions; the associated temperature law is given by~\cite{STbook}
\be
\frac{T_{\rm mU}(t)}{10^9 \mbox{K}}=  \left [ \frac{t (\mbox{yr})}{(\rho/\rho_{\rm nuc})^{1/3}} 
+ \left ( \frac{10^9\,\mbox{K}}{T_0} \right )^6  \right]^{-1/6},
\label{eq:TmU}
\ee
where $T_0 \sim 10^{11}\,\mbox{K}$ is a typical post-merger core temperature  (see e.g.~\cite{foucart16}).
Examples of magnetic field evolution, as described by~\eqref{eq:magfield}, are shown in Fig.~\ref{fig:Bevol}; 
these include a case of static $\ta$ with $B=B_0=3\times10^{17}\,\mbox{G}$ and $T_9=1$ as well as 
two cases of dynamical $\ta$ with $T=T_{\rm mU} (t)$ and $B=(B_0,B_0/2)$ (i.e., this last case considers a four-fold increase in $\ta$). In all cases the $B(t)$ curve remains almost flat 
before its rapid decay at $t \gtrsim \ta$. 
%
%
\begin{figure}
\begin{center}
\includegraphics[width=0.42\textwidth]{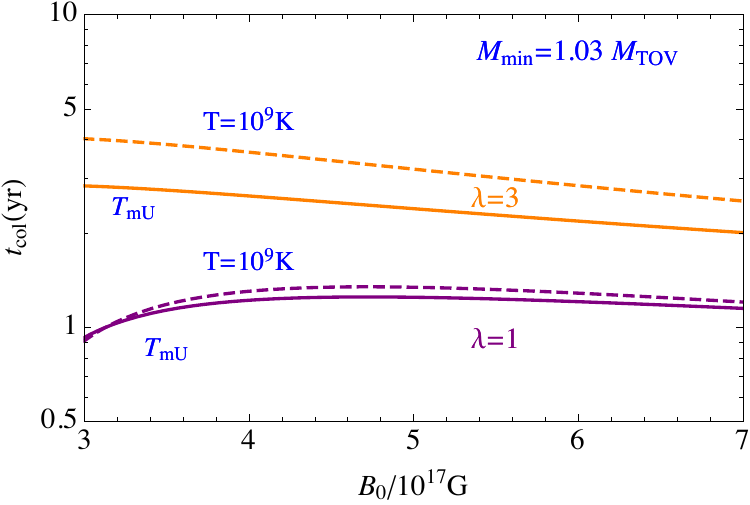}
\end{center}
\caption{The collapse timescale $t_{\rm col}$, calculated from $M_{\rm mag} (t_{\rm col}) = M_{\rm min}=1.03\Mtov$, as a function of the initial 
magnetic field strength $B_0$. The labelled curves represent different choices of the EOS parameter $\lambda$ ($\lambda=1$ and $\lambda=3$ 
for the lower and upper pair respectively) and temperature profile (fixed temperature: dashed; mUrca cooling: solid).}
\label{fig:collapse}
\end{figure}

A reduction of $B$ by a sizable factor should cause the neutron star's prompt collapse. The precise collapse timescale 
$t_{\rm col}$ is defined as the point where the birth mass, 
$M_{\rm min}$, comes to exceed the maximum sustainable by the combined (decaying) magnetic 
and (stable) hydrostatic pressures, i.e. when $M_{\rm mag}(t_{\rm col}) = M_{\rm min}$. Examples of $t_{\rm col} (B_0)$ are shown in Fig.~\ref{fig:collapse} for 
$M_{\rm min} = 1.03 \Mtov$ and different choices of $\lambda$ and temperature evolution $T$ (i.e. static or dynamical $\ta$). 
The curves show marginal variation with $B_0$, which can be taken as evidence of a balance between a faster evolution/larger mass gap 
for higher $B_0$ and a slower evolution/smaller mass gap for lower $B_0$. According to these results, a magnetically supramassive star is unlikely to 
last more than a decade or so after its birth. 

\subsection{Is the collapse inevitable?}
The ambipolar diffusion timescale~\eqref{tamb} assumes normal $npe$ matter without the presence of superconducting protons or 
superfluid neutrons. Ignoring proton pairing is well justified as it is expected to be blocked by  an ultra-strong magnetic field $B_{17} \gtrsim 1$~\cite{baym69}. 
On the other hand, nothing can prevent the onset of neutron superfluidity which is expected to take place at $T_9\approx 0.5-1$ (see e.g.~\cite{acg05,ho12}). 
Once the bulk of the core has undergone the transition to the superfluid state after a time $t=t_{\rm sf}$, the partial decoupling between the charged and 
uncharged fluids is likely to choke ambipolar diffusion and lead to a markedly longer $\ta$~\cite{gjs11}. This implies that the timescale~\eqref{tamb} should 
be accurate for the entire period $t \lesssim t_{\rm sf}$. Assuming~\eqref{eq:TmU},  the aforementioned temperature range translates into a time interval 
$t_{\rm sf} \approx (0.5-33)\,\mbox{yr}$, with the high-end limit being favoured by observations~\cite{page11}.  
{It is worth mentioning that the superfluid phase transition could be further delayed by Pauli-paramagnetic suppression as
a magnetic field $B_{17} \gtrsim 0.1-1$ may prevent the formation of the singlet neutron pairing state, leaving the weaker triplet superfluid state as the only
possibility~\cite{sedrakian17}.}

In fact the assumption of a passively cooling neutron star is not a realistic one; the decaying magnetic field would act as a heat source itself  thus
delaying the onset of superfluidity. The results of Ref.~\cite{ho12} suggest that this delay could be as high as a factor $10-100$, thus making 
$t_{\rm sf} \gg t_{\rm col}$ for most of the parameter space. A similar analysis in \cite{belo16} suggests that ambipolar heating could balance
mUrca cooling at a temperature $T_{\text{bal}} \lesssim 8 \times 10^{8} (B_{16}^2 / L_{5})^{1/5}$ K, likely exceeding the superfluidity-onset 
value until $B$ has sufficiently decayed. We can therefore conclude that neutron superfluidity is unlikely to prevent the 
short-term collapse of magnetically supramassive neutron stars.

Far more serious could be the implications of neutrino cooling via direct Urca reactions~\cite{latt91}. This is classified as a fast cooling mechanism, 
causing the core temperature to plummet down to $T_9 \approx 1$ in a matter of minutes instead of (approximately) the full year required by the 
modified Urca reactions. In such a scenario $t_{\rm sf} \ll t_{\rm col}$, thus preventing an early-stage collapse of the supramassive star.
These direct reactions, however, require that the Fermi momenta of the protons and electrons exceeds that of the neutrons, 
implying a critical proton fraction $x_{\rm p} \gtrsim 0.1$ and an operational density $\rho \gtrsim 4 \rho_{\rm nuc}$~\cite{latt91,belo16}.
Despite their high mass, magnetically supramassive stars may not meet this requirement as a result of their relatively large size compared
to ordinary neutron stars~\cite{card01}.

\section{Observational signatures and closing remarks}
%
Owing to their extreme field strengths, magnetically supramassive stars should be especially active during their relatively short lifetimes. 
SGRBs with extended afterglow, which are thought to be (at least partially) powered by spindown energy injections from a newborn magnetar \cite{rowl13,lg16}, 
are a promising candidate regarding observational signatures. If indeed a magnetically supramassive star was born following a merger, the afterglow `plateau' -- {an often-observed phase of roughly constant X-ray flux \cite{ciolfi20}} -- should be 
short-lived because of intense spin-down, though the luminosity will be exceptionally high since $B$ is large. {Importantly however, if the flux is abruptly truncated, this would indicate a cessation of the magnetar's contribution and a rotationally-supramassive (or accretion-induced) collapse \citep{ravi14}. In the magnetically-supramassive case, the magnetar engine will also eventually shut off, but the collapse should occur sufficiently late such that the drop is undetectable. Generally speaking, after an electromagnetic spin-down timescale, $\tau_{\text{em}} \sim 50 \left( B_{p} / 10^{16} \text{ G} \right)^{-2} \left( \nu / \text{kHz}\right)^{-2}$~s for spin frequency $\nu$, has elapsed, the X-ray luminosity, which traces the spin-down luminosity up to some efficiency factor, would be expected to decay quadratically until becoming invisible due to measurement noise.}   Prototypical examples in this class are 
GRBs 080702A, 100117A, and 100625A, the latter of which seemingly displayed an especially short-lived ($\gtrsim 10$~s) plateau {followed by a power-law decay}, and may have given birth to a magnetar with $B_{p} \gtrsim 10^{17}$~G \cite{rowl13,suvk21}. 

Magnetically supramassive stars would be expected to 
collapse $\gtrsim$ years after birth.
%
Once an event horizon inevitably comes to cloak the star, field lines will snap, inducing magnetic shocks that can accelerate electrons 
to relativistic velocities, producing radiation in the $\gtrsim$ GHz band. This mechanism, though considered only in the context of rotationally supramassive stars,
was put forth 
as a progenitor for extragalactic fast radio bursts (FRBs) \cite{fal14}. {We note that the emitted power in a curvature-radiation scenario scales as $B^2 \nu^2$, and can accommodate the observed FRB energetics even for slow stars (i.e., at times $t \gg \tau_{\text{em}}$) if $B$ is large enough.} Something of a `smoking gun' for magnetically supramassive systems 
may then be a short-lived, bright plateau followed by a power-law decay {(without an abrupt cutoff after $\sim 10^{2}$ s)} after an SGRB, with a (non-repeating) FRB occurring $\gtrsim$ years later once collapse sets in. 
Such a scenario would be difficult to explain with a traditionally supramassive magnetar (since collapse would set in on the much-shorter spindown timescale \cite{ravi14}) 
or a black hole (since fallback accretion would have long since concluded \cite{ciolfi20}).

Late-time X-ray flares are also observed in some afterglow light-curves, sometimes up to $\gtrsim 10^6$ seconds post prompt emission \cite{bern11}.
Even at times $t \ll \ta$, a non-negligible amount of energy may be liberated by ambipolar diffusion if $B$ is extremely large and thus, much like in the case 
of anomalous X-ray pulsars and soft-gamma repeaters \cite{td96}, magnetic dissipation could be responsible for the triggering of these flares. 

Besides their electromagnetic signature, neutron stars with $B_{17} \gtrsim 1$ magnetic fields are expected to 
sustain huge ellipticities (`mountains') $\epsilon$, thus becoming copious sources of GWs with characteristic amplitude 
$h \approx 4 \times 10^{-24} (10\,\mbox{Mpc}/D) (\epsilon/10^{-2}) (\nu / \text{kHz})^{2}$ 
before they have substantially 
spun down \cite{lg16}. Strains of this order may be detectable by aLIGO or next generation detectors, offering another route 
for observational constraints. {The quasi-normal mode spectrum of an ultra-magnetised star would also be significantly shifted relative to that of an otherwise equal but unmagnetised star \cite{lander09}, and thus the ringdown profile could be used as an indicator of whether the system is likely to be magnetically supramassive or not.} {It is also worth noting that to produce a supramassive remnant marginally heavier than $\Mtov$ in a coalescence, it is likely that the pre-merging stars would have to be relatively light, $M \lesssim 1.3 M_{\odot}$, which has implications for GW emission and evolutionary modelling \cite{foucart16}}.

We conclude with a few remarks about future avenues on the modelling of magnetically supramassive stars. 
Based on our earlier discussion, these should include more rigorous, coupled magneto-thermal evolutions that include 
realistic EOS and superfluidity along the lines of Refs.~\cite{ho12,belo16} and the careful delimitation of the dUrca reactions' 
parameter space. In parallel, future GRMHD simulations of coalescing neutron stars with their ever increasing resolution
should be able to provide an improved understanding of the possibility of forming such objects.

\end{document}